\documentclass[]{fairmeta}

\usepackage{hyperref}
\usepackage{url}
\usepackage[utf8]{inputenc} 
\usepackage[T1]{fontenc}    
\usepackage{booktabs}       
\usepackage{makecell}       
\usepackage{amsfonts}       
\usepackage{nicefrac}       
\usepackage{microtype}      
\usepackage{xcolor}         
\usepackage{graphicx}
\usepackage{subcaption}
\usepackage{dsfont}
\usepackage{amsmath}
\usepackage{amssymb}
\usepackage{cleveref}
\usepackage{pifont}
\usepackage[export]{adjustbox}

\usepackage{comment}
\excludecomment{hide}

\usepackage{xcolor}
\newcommand{\model}{TRIBE v2}

\title{A foundation model of vision, audition, and language for in-silico neuroscience}

\author[1]{Stéphane d'Ascoli}
\author[1]{Jérémy Rapin}
\author[1]{Yohann Benchetrit}
\author[1]{Teon Brooks}
\author[1]{Katelyn Begany}
\author[1,2]{Joséphine Raugel}
\author[1]{Hubert Banville}
\author[1]{Jean-Rémi King}

\affiliation[1]{FAIR at Meta}
\affiliation[2]{Laboratoire de Neurosciences Cognitives
et Computationnelles, Ecole Normale Supérieure - PSL}

\abstract{
    Cognitive neuroscience is fragmented into specialized models, each tailored to specific experimental paradigms, hence preventing a unified model of cognition in the human brain.
    Here, we introduce TRIBE v2, a tri-modal (video, audio and language) foundation model capable of predicting human brain activity in a variety of naturalistic and experimental conditions. 
    Leveraging a unified dataset of over 1,000 hours of fMRI across 720 subjects, we demonstrate that our model accurately predicts high-resolution brain responses for novel stimuli, tasks and subjects, superseding traditional linear encoding models, delivering several-fold improvements in accuracy.
    Critically, TRIBE v2 enables in silico experimentation: tested on seminal visual and neuro-linguistic paradigms, it recovers a variety of results established by decades of empirical research. 
    Finally, by extracting interpretable latent features, TRIBE v2 reveals the fine-grained topography of multisensory integration. 
    These results establish artificial intelligence as a unifying framework for exploring the functional organization of the human brain.

}

\date{\today}
\correspondence{\email{sdascoli@meta.com} and \email{jeanremi@meta.com}}

\metadata[Code]{\url{https://github.com/facebookresearch/tribev2}}
\metadata[Weights]{\url{https://huggingface.co/facebook/tribev2}}
\metadata[Demo]{\url{https://aidemos.atmeta.com/tribev2}}

\begin{document}

\maketitle

\begin{figure*}[htb]
    \centering
    \includegraphics[width=.95\linewidth]{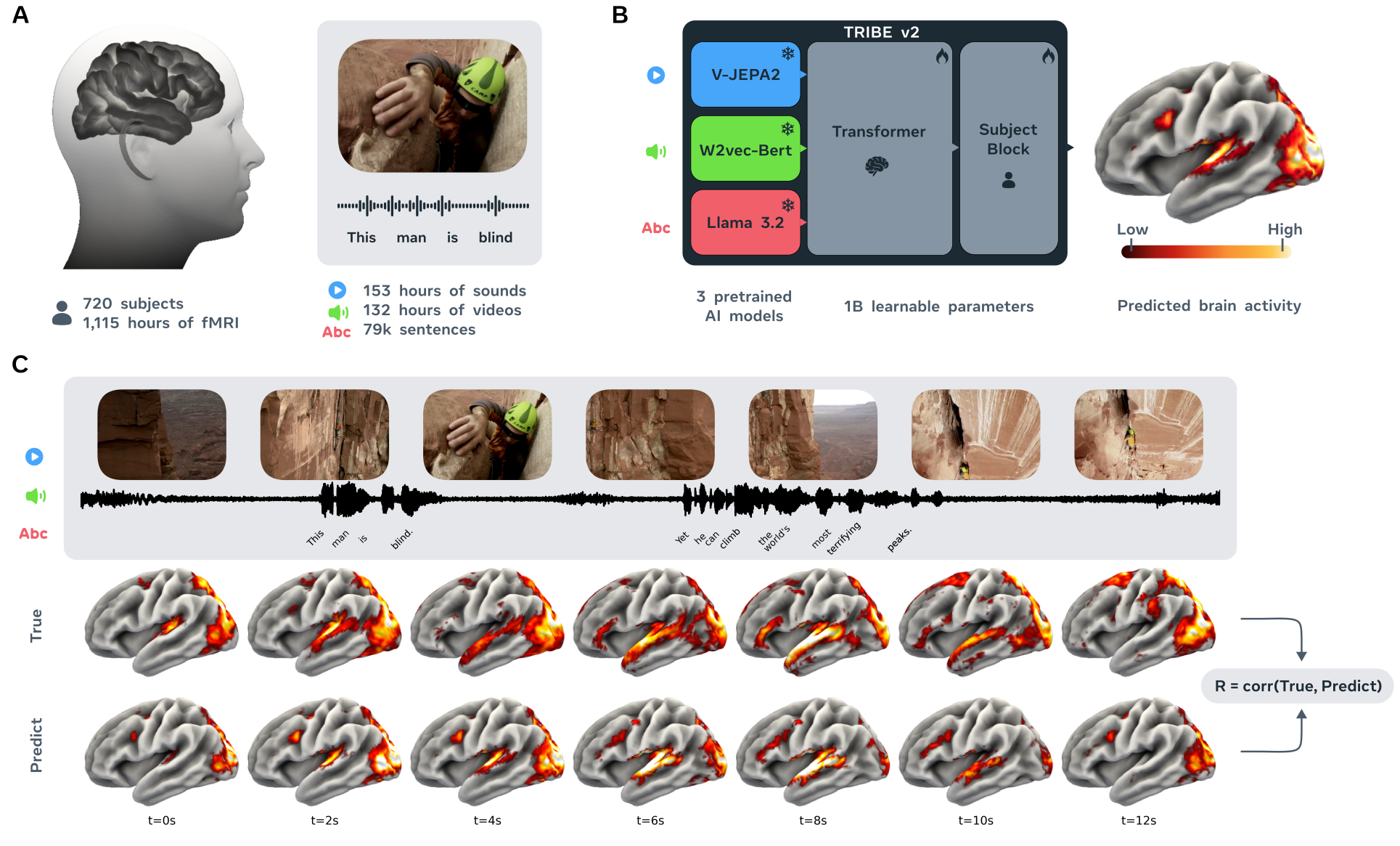}
    \caption{\textbf{Approach. A.} 
    The brain activity of healthy volunteers was recorded with functional Magnetic Resonance Imaging (fMRI), while they were presented to naturalistic (movies, podcasts) and experimental conditions (e.g. flashed objects, isolated words). 
    \textbf{B.} \model{} is trained to predict high-resolution fMRI from any audio, video, and/or text stimuli, using embeddings from pretrained AI models~\cite{grattafiori2024llama,chung2021w2v,assran2025v}. \textbf{C.} Example of the average brain response (mean across 176 subjects) to a movie clip~\citep{van2013wu}, and the corresponding zero-shot prediction from \model{}.
    }
    \label{fig:summary}
\end{figure*}

\section{Introduction}

Neuroscience has historically progressed through a highly specialized study of cognitive functions and their underlying neural substrates. In the domain of vision, for instance, research has systematically mapped the representations of motion to area V5~\citep{shadlen2001neural}, faces to the fusiform gyrus~\citep{kanwisher2006fusiform}, and written language to the visual word-form area~\citep{dehaene2011unique}. While this "divide-and-conquer" approach has yielded deep insights, the resulting landscape remains fragmented and difficult to synthesize. Understanding how neuronal assemblies represent and integrate information to form a coherent model of the surrounding world remains a fundamental challenge ~\citep{mathis2024decoding,yamins2016using,richards2019deep}. 

The fast progress of foundational models for language, vision, and audio offers a promising resolution to this challenge.
Indeed, increasing evidence suggests that the latent hierarchies of deep neural networks exhibit a striking convergence with the representational geometry of the primate brain~\citep{kay2008identifying, yamins2014performance, van2017artificial, kriegeskorte2015deep}. This alignment effectively enables direct prediction of brain responses to natural images~\citep{yang2023memory,Adeli2023.08.02.551743,nguyen2023algonauts,ozcelik2023natural,scotti2024mindeye2}, videos \citep{lahner2025mosaic,sartzetaki2024one}, sound~\citep{kell2018task,millet2022toward,giordano2023intermediate} and text~\citep{mitchell2008predicting,huth2016natural,toneva2019interpreting,schrimpf2018brain,caucheteux2022brains} by a linear transformation of the neural network's latent representations.

The representational alignment between brains and algorithms delineates a path toward a foundation model of human brain function -- derived not from first principles, but from the direct mapping of large amounts of brain responses to pretrained AI architectures. For this goal to be successful, four essential criteria must be met: 
first, \textbf{integration}, where the model captures whole-brain responses across a vast repertoire of experimental conditions; 
second, \textbf{performance}, reaching or exceeding the predictive accuracy of traditional analytical pipelines tailored to specific neural response patterns; 
third, \textbf{generalization}, allowing the model to ``zero-shot'' generalize to novel experimental conditions and thus equip researchers with a tool to improve the design, data-efficiency, and analysis of their experiments; 
and finally, 
\textbf{interpretability}, providing a mechanistic toolkit to decompose the organization of cognitive functions and neural representations.

Here, we present \textbf{\model{}}, a tri-modal (audio, video, and language) foundation model for human brain activity. Building on the v1 architecture~\citep{d2025tribe} -- which achieved state-of-the-art performance in the 2025 Algonauts challenge \citep{gifford2024algonauts, st2023cneuromod} -- we here scale this approach to high-resolution fMRI and evaluate it on a variety of "deep-" and "wide-" datasets, together encompassing over 1,000 hours of fMRI recordings across 720 subjects \cref{tab:datasets_comparison}. We demonstrate that \model{}: (1) accurately predicts cortical responses across diverse naturalistic and experimental conditions; (2) enables rapid, \textit{in silico} hypothesis testing; and (3) provides a unified framework to accelerate neuroscientific discovery.

\begin{table*}[ht]
    \centering
    \scriptsize
    \begin{tabular}{l|cccccccccc}
    \toprule
    \textbf{Dataset} & \textbf{Mode} & \textbf{Modalities} & \textbf{Device} & \textbf{\makecell{Subjects}} & \textbf{\makecell{Sessions}}  & \textbf{\makecell{fMRI\\(h)}} & \textbf{\makecell{Video\\(h)}} & \textbf{\makecell{Audio\\(h)}} & \textbf{\makecell{Sentences}}\\
    \midrule
CNeuroMod~\citep{st2023cneuromod} & Train & A+V+T & 3T & 4 & 1408 & 268.7 & 64.5 & 64.5 & 54k\\
BoldMoments~\citep{lahner2024modeling} & Train & A+V & 3T & 10 & 520 & 61.9 & 33.2 & 32.2 & -\\
Lebel2023~\citep{lebel2023natural} & Train & A+T & 3T & 8 & 432 & 85.8 & - & 16.1 & 5k\\
Wen2017~\citep{wen2018neural} & Train & V & 3T & 3 & 258 & 35.2 & 3.1 & - & -\\
\midrule
Subtotal & Train &  &  & 25 & 2618 & 451.6 & 100.7 & 112.7 & 59k\\
\midrule
NNDb~\citep{aliko2020naturalistic} & Test & A+V+T & 3T & 86 & 86 & 160.6 & 19.4 & 19.4 & 3k\\
LPP~\citep{li2022petit} & Test & A+T & 3T & 112 & 1008 & 180.2 & - & 4.8 & 4k\\
Narratives~\citep{nastase2021narratives} & Test & A+T & 3T & 321 & 678 & 146.6 & - & 4.4 & 4k\\
HCP~\citep{van2013wu} & Test & A+V+T & 7T & 176 & 704 & 178.7 & 1.0 & 1.0 & 496\\
\midrule
Subtotal & Test &  &  & 695 & 2476 & 666.1 & 20.5 & 29.7 & 11k\\
\midrule
Total & All &  &  & 720 & 5094 & 1117.7 & 121.1 & 142.4 & 71k\\
    \bottomrule
    \end{tabular}
    \caption{\textbf{Characteristics of the datasets}. Wide datasets prioritize population-level scale and generalizability, whereas deep datasets prioritize individual-level granularity and precision.}
    \label{tab:datasets_comparison}
\end{table*}

    
\begin{figure*}[ht]
    \includegraphics[width=\linewidth]{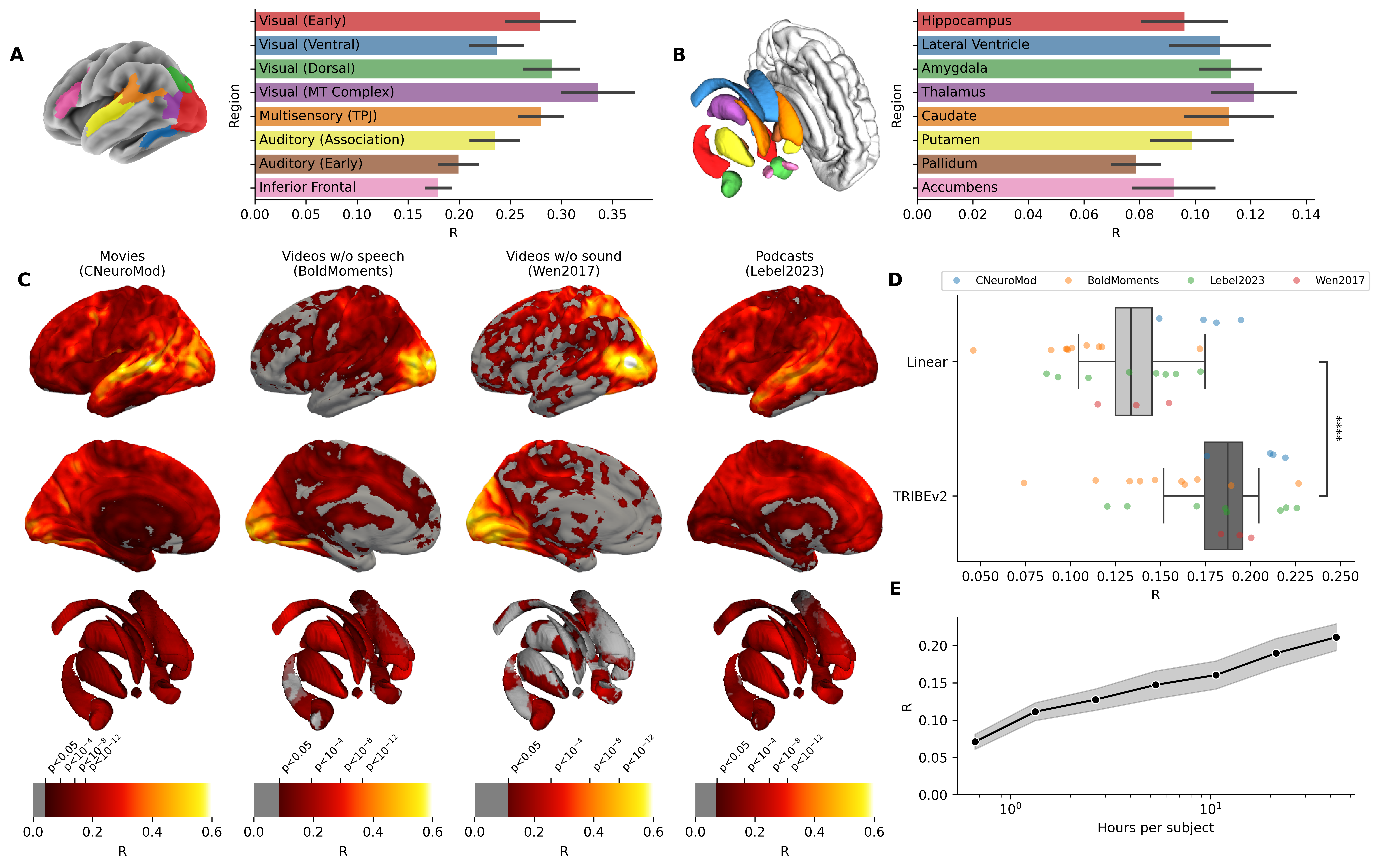}
\caption{\textbf{\model{} predicts fMRI responses accurately across the whole brain.}
\textbf{A.} Encoding scores in various important regions of the cortex as defined by the HCP parcellation (see App.~\cref{fig:rois} for more details), color-coded in the cortical surface on the left. Error bars denote Standard Error of the Mean (SEM) across the participants of all studies.
\textbf{B.} Encoding scores in the 8 subcortical regions considered, color-coded in the figure on the left. Error bars denote SEM across the participants of all studies.
\textbf{C.} Subject-averaged encoding scores across cortical and subcortical regions for the four conditions in the training dataset. 
\textbf{D.} Average encoding scores of individual subject (dots) for \model{} and a classic linear baseline.
\textbf{E.} Scaling laws of brain encoding. Average encoding score of \model{} on the Algonauts dataset, as we increase the number of hours of training data used per participant. Error bar indicates the SEM across subjects.
}
\label{fig:performance}
\end{figure*}

\begin{figure*}[htb]
    \includegraphics[width=\linewidth]{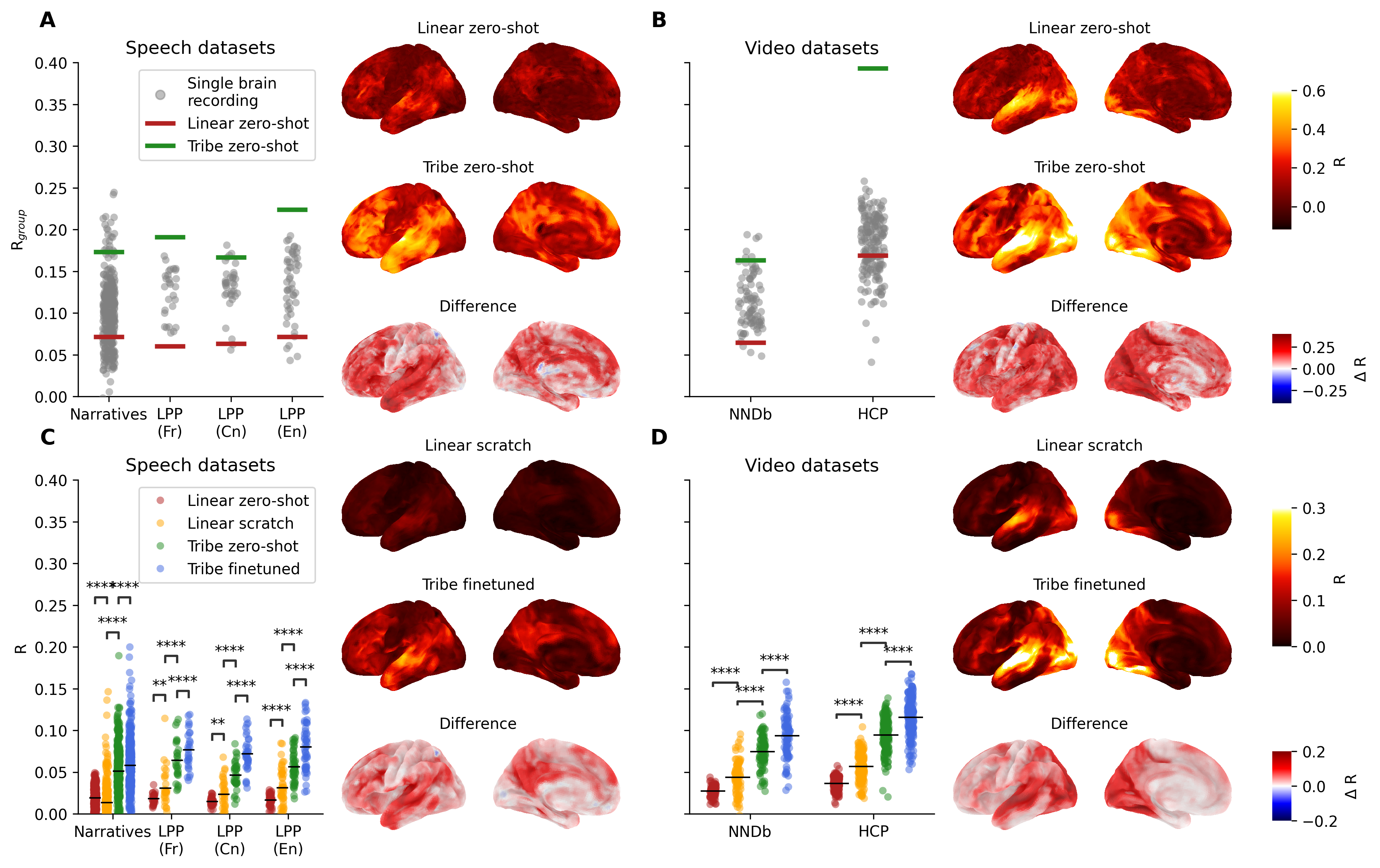}
    \caption{\textbf{\model{} generalizes zero-shot to new tasks and subjects, and can be finetuned on a small amount of data to improve individualized predictions.} 
    \textbf{A}. Left: Pearson correlation between (i) the brain response to speech of an individual subject $k$ and (ii) the average brain response of the rest of the cohort. Each dot represents a subject. The bars represent the correlation between the average cohort response and the zero-shot predictions from \model{} (green) and its linear counterpart (red). Right: cortical map of the scores achieved, averaged across datasets. \textbf{B}. Same as A but for videos instead of speech.
    \textbf{C}. Left: encoding scores of \model{} for all subjects of the unseen speech datasets (each dot represents a subject and the horizontal lines represent the mean). We compare the results obtained with a linear model (either zero-shot or trained from scratch) to those obtained with \model{} (either zero-shot or finetuned on half of the unseen subjects' data). Right: cortical map of the scores achieved, averaged across datasets. \textbf{D}. Same as C but for videos instead of speech.
    \\
    }
    \label{fig:ood}
\end{figure*}



\begin{figure*}
    \centering
    \includegraphics[width=\linewidth]{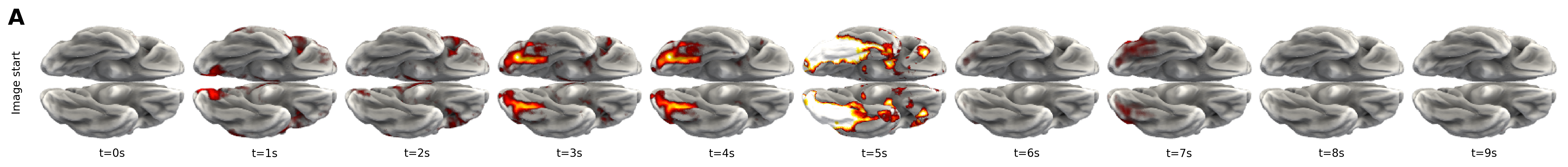}
    \includegraphics[width=\linewidth]{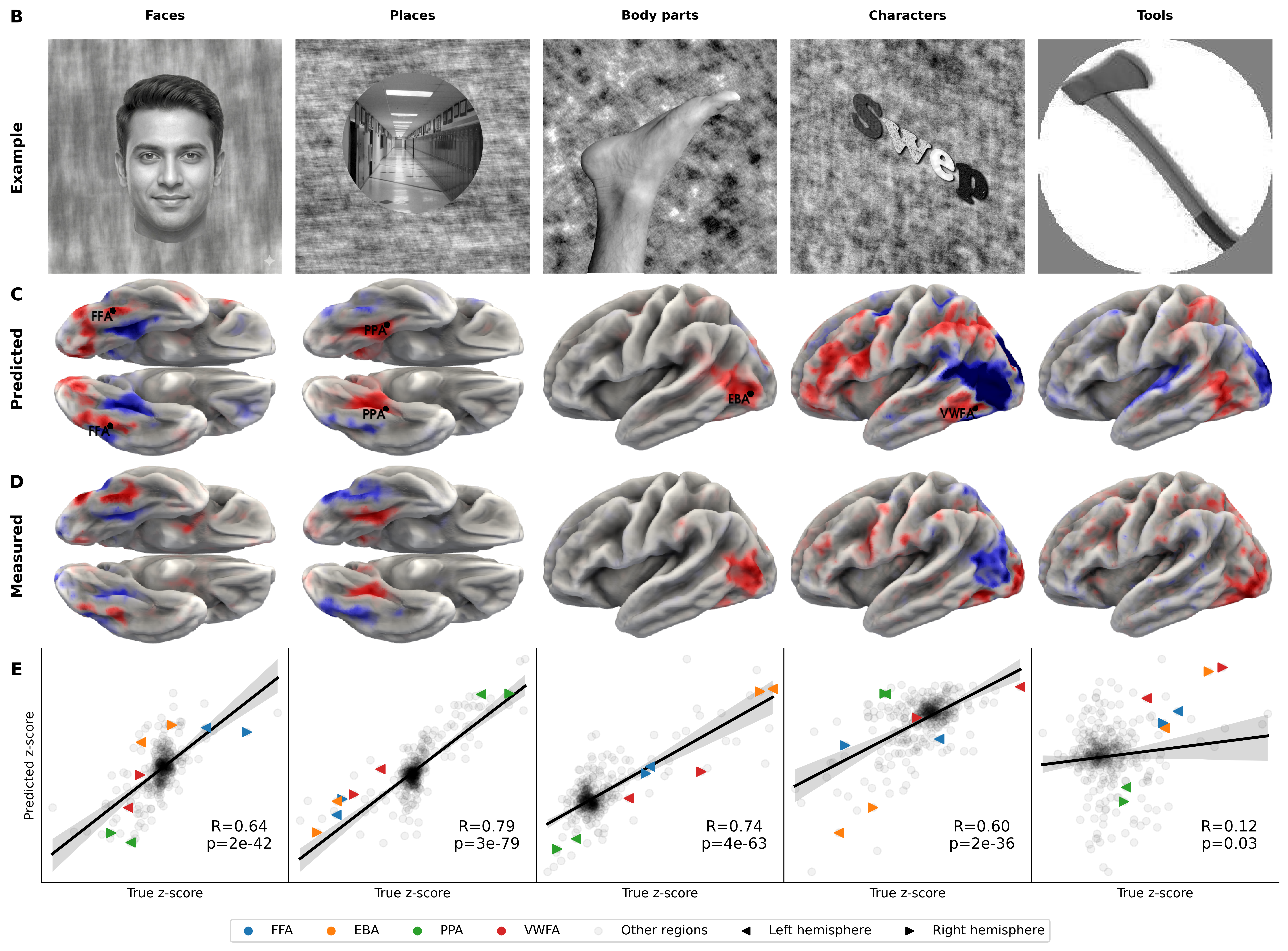}
    \caption{\textbf{\model{} recovers, in-silico, a variety of findings in visual neuroscience.}
    \model{} is tested on various visual functional localizers, each extracted from the Individual Brain Charting (IBC) dataset~\citep{pinho2018individual}. Experimental images are flashed for 1 second, with an interval of 8 seconds, and a general linear model (GLM) is fit on the predicted time-series to obtain $z$-scored contrast maps (see \cref{methods:in_silico} for details).
    \textbf{A.} Mean evoked response to the images of the \textit{FaceBody} task.
    \textbf{B.} Examples of images from each category. The images are taken from the \textit{FaceBody} and \textit{Visu} tasks from the IBC dataset.
    \textbf{C.} Contrast maps obtained from TRIBE's predictions (see \cref{methods:in_silico}).
    \textbf{D.} Contrast maps computed on the IBC dataset, averaged across participants and trials.
    \textbf{E.} Numerical agreement between ground truth and predicted z-scores. Each dot represents one of the 360 parcels from the HCP parcellation~\citep{glasser2016multi}, and z-scores are averaged for all cortical vertices falling within the parcel. Functionally relevant parcels are indicated in colors, with the marker disambiguating the hemisphere. Higher correlations between true and predicted brain activity patterns indicate better zero-shot in-silico experiments.
    }
    \label{fig:in_silico_vision}
\end{figure*}

\begin{figure*}
    \centering
    \includegraphics[width=\linewidth]{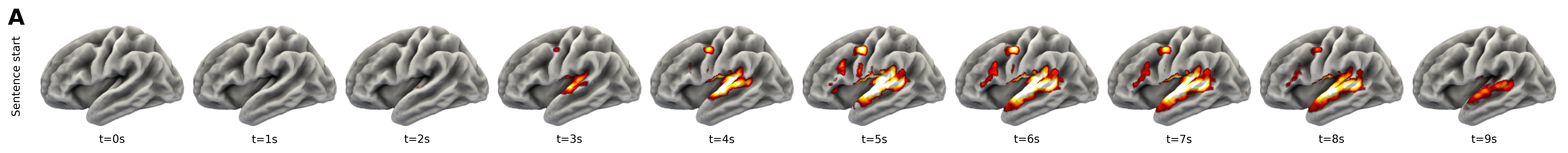}
    \includegraphics[width=\linewidth]{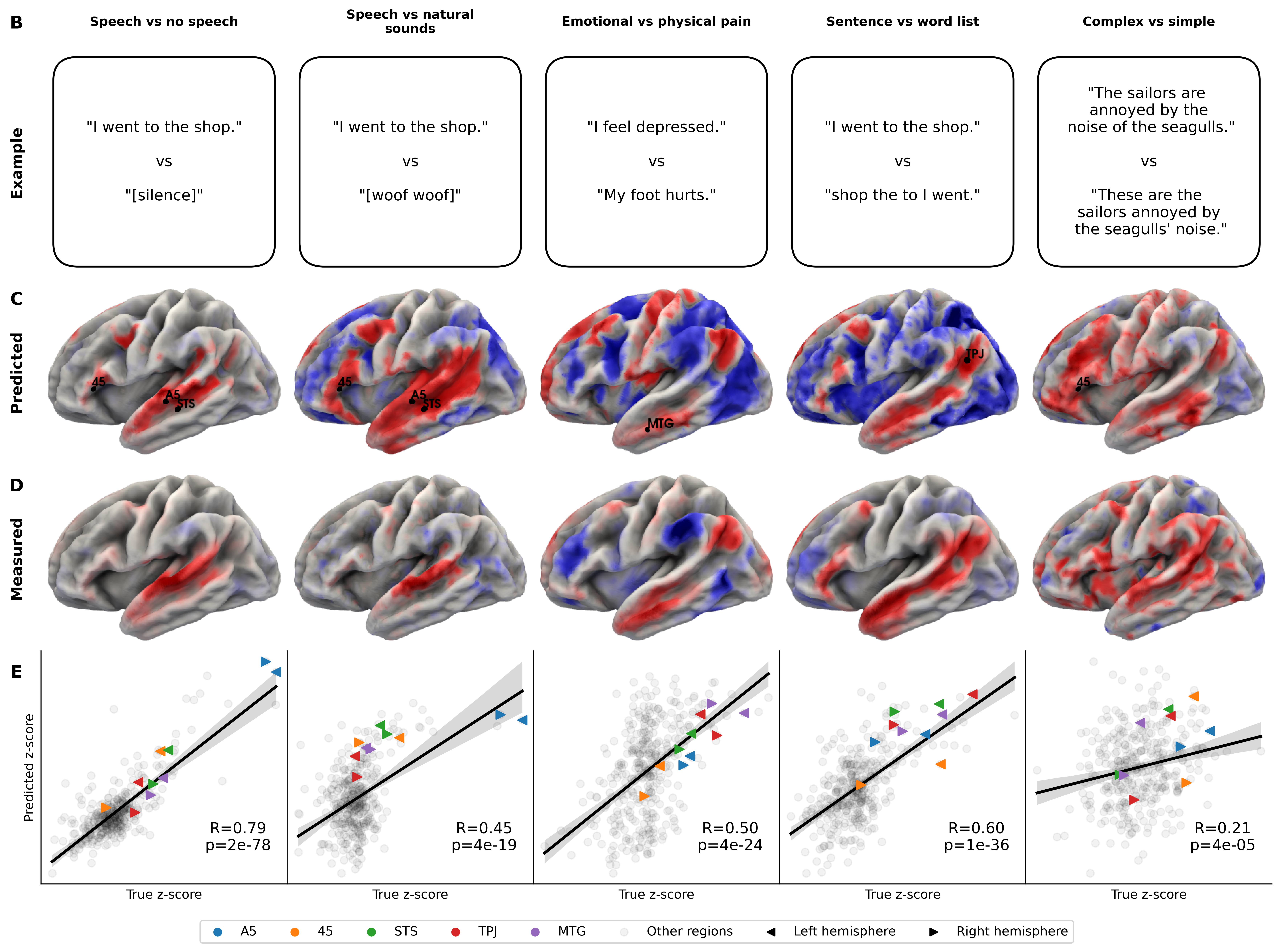}
    \caption{\textbf{\model{} recovers, in-silico, a variety of classic findings in neurolinguistics.}
    \model{} is tested with various language functional localizer tasks, extracted from the Individual Brain Charting (IBC) dataset \citep{pinho2018individual}.
    \textbf{A.} Mean evoked response prediction to ten-word sentences in the "simple" condition of the \textit{RSVP} task.
    \textbf{B.} In-silico experiment categories. These conditions are extracted from the \textit{Bang}, \textit{ArchiSocial}, \textit{EmotionalPain} and \textit{RSVP} tasks from IBC.
    \textbf{C.} Contrast maps obtained from TRIBE's predictions (see \cref{methods:in_silico}).
    \textbf{D.} Contrast maps computed on the IBC dataset, averaged across participants and trials.
    \textbf{E.} Numerical agreement between ground truth and predicted z-scores. Each dot represents one of the 360 parcels from the HCP parcellation~\citep{glasser2016multi}, and z-scores are averaged for all cortical vertices falling within the parcel. Functionally relevant parcels are indicated in colors, with the marker disambiguating the hemisphere.
    }
    \label{fig:in_silico_language}
\end{figure*}

\begin{figure*}
    \includegraphics[width=\linewidth]{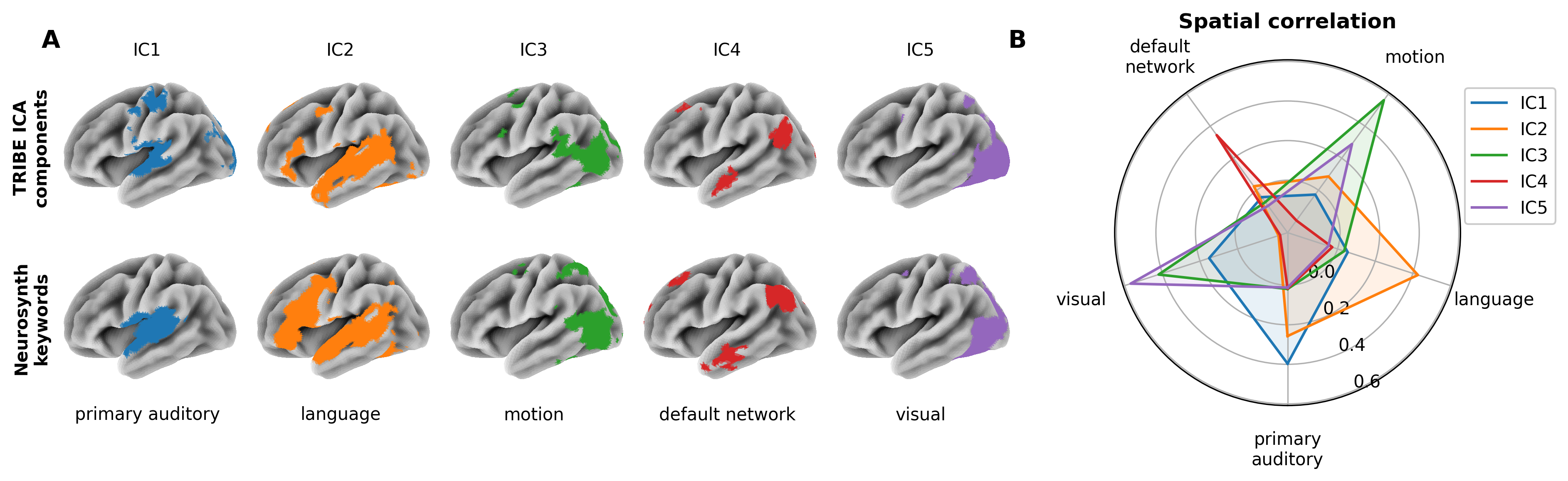}
    \caption{\textbf{Independent component analysis (ICA) shows that \model{} learns neuroscientifically relevant patterns.} 
    \textbf{A.} Top. \model{}'s ICA. Bottom semantic maps for five neuroscientific keywords generated by Neurosynth \citep{kent2026neurosynth}. In both cases, the 10\% vertices with the highest values are shown.
    \textbf{B.} Spatial correlation (R) between each ICA component and the functional networks to well-known functional networks.
    }
    \label{fig:ica}
\end{figure*}

\begin{figure*}[htb]
    \centering
    \includegraphics[width=\linewidth]{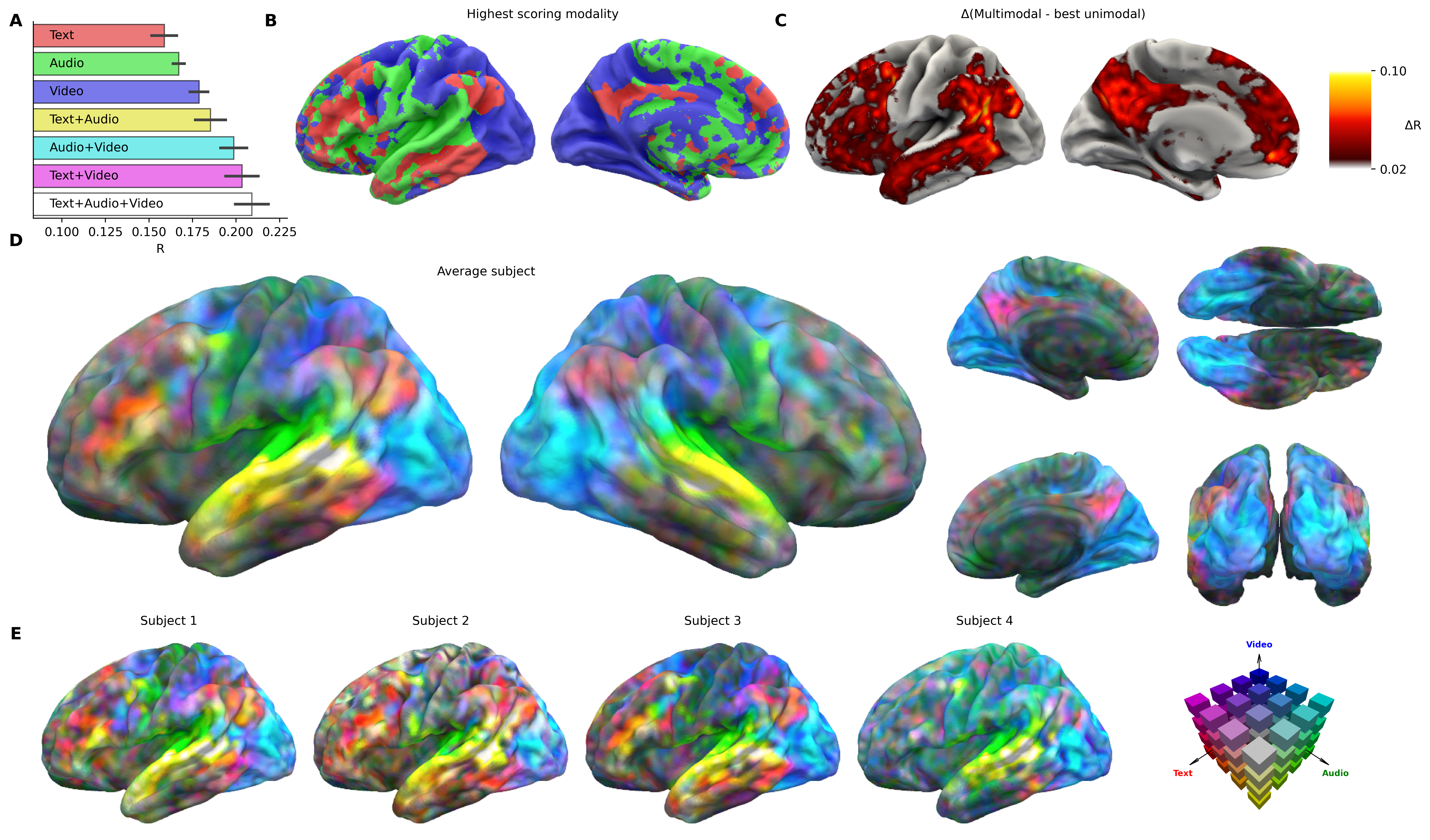}
    \caption{
    \textbf{\model{} reveals insights into the multimodality of the human brain.}
    \textbf{A.} Average cortical encoding scores of \model{} when trained on subsets of the three modalities for the Algonauts dataset. Error bars denote SEM across the four participants.
    \textbf{B.} Highest scoring modality across the whole cortex, with red, green and blue depicting text, audio and video respectively.
    \textbf{C.} Difference between the encoding score of \model{} and the best of its unimodal counterparts.
    \textbf{D.} Each vertex of the cortical and subcortical sufaces are color-coded using an RGB mapping where red, green and blue intensities are determined by the subject-averaged encoding score of the text, audio and video unimodal versions of \model{} on the Algonauts dataset. \textcolor{red}{Red}, \textcolor{blue}{blue} and \textcolor{green}{green} areas correspond to unimodal areas well encoded by text, audio and video respectively, while \textcolor{magenta}{magenta}, \textcolor{yellow}{yellow} and \textcolor{cyan}{cyan} correspond to bimodal areas well encoded by text+video, text+audio and video+audio respectively. Saturation is normalized to make  colors more readable.
    \textbf{E.} Same as D for each of the four subjects in the Algonauts dataset.
    }
    \label{fig:multimodal}
\end{figure*}

\section{Results}

\subsection{Encoding performance across naturalistic tasks}


To evaluate \model{}, 
we first assess its ability to reliably predict the brain responses to naturalistic stimuli on subjects used during training. Specifically, we hold out fMRI responses to naturalistic conditions and correlate them with \model{}'s predictions given video, sound and texts (\cref{methods:model}). 
The results show that a wide variety of cortical and subcortical regions are predicted above chance (\cref{fig:performance}A-C; see \cref{methods:stats} for details on statistical tests).
As expected, the spatial distribution of these predictions vary with the task: podcast listening leads to peak correlation in the temporal cortices (Lebel2023); video watching (Wen2017 and BoldMoments) leads to peak correlations in the visual cortices; and multimodal stimuli (Courtois NeuroMod) combines these two peaks, yielding statistically significant predictions across most of the cortex. 
These relatively high cortical scores contrast with those achieved in subcortical regions, which tend to be more uniform and lower by two to three folds, but remain significant in most areas.
Overall, these predictive performance scores suggest that \model{} adequately captures a vast repertoire of naturalistic conditions.

\subsection{Comparison to baselines}

To contextualize the performance of \model{}, we then compare it against a high-capacity implementation of the Finite Impulse Response (FIR) model~\citep{dale1999optimal} -- the current gold standard for voxel-wise encoding. 
Traditional FIR models are typically constrained to single-subject analysis, hand-crafted features and small-scale datasets. To provide a more rigorous baseline, we here developed a "Deep FIR" encoder optimized via stochastic gradient descent and fed with the same state-of-the-art pretrained embeddings used by \model{} (\cref{methods:model}). This ensures that the performance gap reflects the architectural advantage of our transformer-based integration rather than differences in input features or optimization scale. 
Across all datasets, \model{} significantly outperforming the optimized linear baseline ($q(FDR) < 10^{-4}$, t-test across subjects), demonstrating the advantage of deep nonlinear methods to model brain responses (\cref{fig:performance}D).
We further characterize the scalability of our architecture by analyzing performance as a function of data volume. We observe a log-linear increase in encoding accuracy across the Courtois NeuroMod dataset without any plateau (\cref{fig:performance}E), suggesting that \model{} is uniquely positioned to benefit from the ongoing expansion of neuroimaging repositories.
We first showed this scaling capability in the Algonauts 2025 competition, where an earlier iteration of our \model{} architecture achieved first place out of 263 teams, establishing a new state-of-the-art for the field (\cref{tab:leaderboard}).

\subsection{Generalization to new subjects}

To assess whether \model{} can generalize to new subjects, we curated four naturalistic studies for testing consisting of many subjects recorded for a short amount of time (\cref{tab:datasets_comparison}). We first evaluate the ability of \model{} to predict responses without any retraining, by computing the correlation $R_{group}$ between its predictions and the group-averaged response (see \cref{methods:model} for implementation details). We compare this group-level correlation with the ``group-predictivity'' of subject $i$, i.e. the correlation between subject $i$'s response and the group response. Strikingly, across all datasets considered, \model{}'s predictions provide a more accurate estimation of the group response than the recording of most individual subjects (\cref{fig:ood}A-B). 
The gap is most notable in the Human Connectome Project (HCP) dataset, which features the best signal-to-noise ratio (it is the only dataset recorded with a 7T scanner, and features the most repetitions of each stimulus): here, \model{} boasts an $R_{group}$ near 0.4, a two-fold improvement over the median subject's group-predictivity. 

\subsection{Finetuning to improve individual brain modeling}

While these results confirm that \model{} can model subject-averaged responses of unseen studies, it remains to be tested whether it can model subject specificities with the limited data per participant available in these broad studies. For this, we hold out half of the data for each participant (at most one hour) and finetune \model{} for one epoch on each subject (see \cref{methods:training}). This significantly increases the encoding scores in all datasets considered ($q(FDR)<10^{-4}$, t-test across subjects), and leads to a two- to four-fold improvement over a linear encoder trained from scratch on each subject (\cref{fig:ood}C-D).

Overall, these results demonstrate that \model{} can be used to predict brain responses for unseen participants, either by modelling the group-averaged response in a zero-shot manner or by using a small amount of fine-tuning data to model subject specificities. Crucially, the fact that \model{} predicts group-averaged responses better than individual responses suggests that it could readily be used for piloting naturalistic studies. However, how would such performance hold for non-naturalistic studies?

\subsection{In-silico experiments: vision}

While the studies considered here all involve naturalistic perception tasks, the vast majority of neuroscientific experiments involves controlled stimuli, aimed at isolating a specific phenomenon in the brain. An example for this is the Individual Brain Charting (IBC) dataset~\citep{pinho2018individual}, which consists in a battery of classic experiments designed to obtain functional localizers, e.g. to identify the brain areas involved in face recognition or language processing. To test whether \model{} is reliable in such controlled settings, we replicate some of these experiments \textit{in silico} to compare its predictions with actual results achieved by averaging the responses from participants of the IBC dataset. 

We begin by replicating classic visual functional localizers from IBC: various categories of images~(\cref{fig:in_silico_vision}B) are flashed for 1 second every 8 seconds. Note that although this type of protocol is far from naturalistic, \model{} yields the expected dynamical response (\cref{fig:in_silico_vision}A): activity increases in the ventral visual stream, and peaks 5 seconds after the stimulus onset, as expected from the hemodynamical delay. 
Various contrast maps are then computed following the IBC protocol, as detailed in~\cref{methods:in_silico}. A qualitative match is observed between the areas predicted by \model{} (\cref{fig:in_silico_vision}C) and those measured in the original experiments (\cref{fig:in_silico_vision}D), which is quantified by a significant spatial correlation between the corresponding maps (\cref{fig:in_silico_vision}E). In particular, well-known responsive areas are recovered: the fusiform face area (FFA) for faces, the parahippocampal place area (PPA) for places, the extrastriate body area (EBA) for bodies and the visual word-form area (VWFA) for written characters (\cref{fig:in_silico_vision}E). 

\subsection{In-silico experiments: language}

Next, we replicate language experiments from the IBC dataset. For this, we convert the textual stimuli to an audio file from which we extract word timings, and feed both the text and audio embeddings to \model{}. As shown in \cref{fig:in_silico_language}A, \model{}
yields the expected response to short (10-word) sentences: the elicited response begins at t=3 seconds in the primary auditory cortex then propagates to the whole language network. 
We begin with two localizers of the language network, which contrasts speech segments to either non-speech segments of a movie or natural sounds (\cref{fig:in_silico_language}B). This selects the core language processing areas (\cref{fig:in_silico_language}C-E), in particular associative auditory cortices (A5) and to a lesser degree the superior temporal sulcus (STS) and Broca's area (45). 

Next, we test the ability of \model{} to isolate emotional processing regions by contrasting sentences relative to emotional versus physical pain. \model{} correctly recovers two important regions: the temporo-parietal junction (TPJ) and the middle temporal gyrus (MTG). Finally, we replicate two common linguistic contrasts: sentences versus word lists and complex versus simple sentences. In these two settings, \model{} correctly predicts a higher lateralization to the left hemisphere, as well as higher responses in semantic regions (e.g. TPJ) for the former and syntactic regions (e.g. Broca) for the latter. 

Overall, the striking qualitative agreement between the predictions of \model{} and the outcomes of classic experiments from the literature demonstrate that beyond its encoding performance in naturalistic datasets, \model{} holds promise for the actual downstream task which matters for neuroscience: \textit{in silico} experimentation~\citep{jain2024computational,gifford2025silico}.


\subsection{\model{} learns interpretable representations}
 
To what extent does \model{} learn interpretable representations of the brain? 
To investigate this question, we apply an Independent Component Analysis (ICA) to the final layer of the model, which maps the latent space of the model to the cortical space (details in \cref{methods:ica}). Strikingly, the top five components closely resemble five well-studied function networks in neuroscience~(\cref{fig:ica}A): the primary auditory cortex, the language network, the motion detection area, the default mode network and the visual system \citep{damoiseaux2006consistent}. This is confirmed by comparison with cortical maps obtained via NeuroSynth, a meta-analysis tool which obtains functional maps from common keywords in the neuroscientific literature~\citep{kent2026neurosynth}: each component has a high spatial correlation with a single of these functional networks (\cref{fig:ica}B).

\subsection{Insights into multimodality}

To what extent do the three modalities combined by \model{} contribute to encoding performance? We address this question in \cref{fig:multimodal}A, by assessing the encoding performance of \model{} retrained with various modalities ablated. When training on a single modality, \model{} achieves significantly lower encoding scores: text achieves the lowest average encoding score overall, followed by audio, then video. These modalities encode complementary parts of the brain (\cref{fig:multimodal}B): as expected, audio dominates near the auditory cortex, video dominates in the occipital and partietal cortices, while text, which presumably contains the most semantic information, dominates in the language processing cortices but also in large parts of the prefrontal lobe.

In which areas does multimodality yield the strongest gains? In \cref{fig:multimodal}C, we compare for each parcel the encoding score of the multimodal encoder with that of the best of the three unimodal encoders. We observe that the largest gains are around the temporal-parietal-occipital junction (with up to 50\% increase in encoding score), and to a lesser degree in the prefrontal cortex. 
These results demonstrate that our multimodal encoder effectively captures interactions between modalities, which improves whole-brain encoding.

To achieve a more precise understanding of multisensory integration, we overlay the contribution of the three modalities using an RGB encoding where red, green and blue respectively represent the encoding scores achieved solely with text, audio and video (\cref{fig:multimodal}D). The mixing between colors quantifies multisensory integration: for example, text+audio (yellow) can be observed in the superior temporal lobe and parts of the ventricles, and video+audio (cyan) can be observed in the ventral and dorsal visual cortices, as well as in the hippocampus.
Note that different subjects display subtle but meaningful differences in the spatial organization of the multimodality (\cref{fig:multimodal}E). 

Overall, these observations not only align qualitatively with what one would expect from a neuroscientific point of view (e.g. \citet{stein2008multisensory,driver2008multisensory,gao2023audiovisual,tang2023brain,hu2025neural}), but also yield valuable insights on how multisensory integration may occur in the human cortex.

\begin{hide}
\end{hide}

\section{Discussion}
The present results strengthen the possibility of a paradigm shift in neuroscience \citep{hamilton2020revolution}, moving from the fragmented mapping of isolated cognitive tasks toward the use of unified, predictive foundation models of brain and cognitive functions \citep{mensch2021extracting,schneider2023learnable,azabou2023unified,caro2023brainlm,wang2025foundation,Binz2025Centaur}. By aligning the representations of AI systems to those of the human brain, we demonstrate that a single architecture can integrate a vast range of fMRI responses across hundreds of individuals, extending the framework that led the 2025 Algonauts competition \citep{d2025tribe}. The observed log-linear scaling of encoding accuracy—mirroring power laws in both artificial intelligence \citep{kaplan2020scaling} and neuroscience \citep{antonello2023scaling,beliy2024wisdom,banville2025scaling,dascoli2025decoding} -- suggests that the ceiling for predicting human brain activity is yet to be reached. Unlike traditional alignment methods such as hyperalignment \citep{haxby2011common}, which require participants to be exposed to shared stimuli, \model{}’s ability to generalize to novel subjects and unseen protocols (\cref{fig:ood}) establishes an \textit{in silico} platform for pre-screening neuroimaging protocols and augmenting the statistical power of existing datasets. Similarly, and unlike recent \textit{decoding} models that reconstruct words ~\citep{tang2023semantic,defossez2023decoding} or images from brain activity~\citep{ozcelik2023natural,benchetrit2023brain,scotti2024mindeye2}, \model{} serves as a general-purpose \textit{encoding} model of whole-brain activity across cognitive domains.

Despite its predictive breadth, \model{} remains constrained by the inherent spatio-temporal resolution of fMRI, which cannot capture the millisecond dynamics of neuronal firing. Furthermore, while the model recovers several canonical neural responses — including selectivity for places and faces \citep{kanwisher2006fusiform,chang2017code} as well as linguistic functions~\citep{pallier2011cortical,dehaene2011unique,friederici2017language,fedorenko2024language,bhaya2022speech} — its current inputs are limited to visual, auditory, and semantic features, and thus omit primary sensory modalities such as olfaction, balance, somatosensation, and their integration \citep{hedger2025vicarious}. A more fundamental limitation is that the model currently treats the brain as a passive observer of naturalistic stimuli; it does not yet model the brain as an active agent producing behavior. Integrating neuro-developmental trajectories and clinical pathology remains a primary goal to move beyond a static, adult brain state and capture the full diversity of the global population \citep{henrich2010weirdest}.

The transition toward foundation models in neuroscience follows a trajectory recently established in structural biology \citep{jumper2020alphafold}, chemistry \citep{abed2024open} and other domains of neuroscience \citep{wang2025foundation}, where data-driven unification replaces a series of isolated, small-scale observations. While the "black box" nature of deep neural networks can obscure the exact premises of their predictions \citep{linardatos2020explainable}, \model{} demonstrates that these models can nonetheless decompose cortical functional hierarchies and anchor them to specific neural representations \citep{huth2016natural,margulies2016situating}. As a robust "digital model" of the human brain capable of recovering the results of both deep-phenotypic datasets like the Individual Brain Charting (IBC) repository \citep{pinho2018individual} and large-phenotypic ones like the Human Connectome Project~\citep{van2013wu}, \model{} provides a platform for interpreting neural function through intervention. As these models continue to scale with the accumulation of naturalistic datasets \citep{van2013wu,poldrack2017scanning,pinho2018individual,hebart2019things,nastase2021narratives,allen2022massive,st2023cneuromod,lebel2023natural}, they will become indispensable tools for testing existing theories \citep{poldrack2017scanning} and identifying the specific experiments most likely to improve our understanding of brain mechanics.

\newpage
\section{Acknowledgements}
The authors wish to thank Fernanda Ponce and Bertrand Thirion for their help with the IBC data \citep{pinho2018individual}; Pierre-Louis Xech, Elisa Cascardi and Jennifer Pak for their support; Diego Marcos Segura and Dominic Giardini for the demo webpage; Valentin Wyart for his advices, and the rest of the Brain and AI team at Meta AI for insightful discussions.

The Courtois project on neural modelling was made possible by a generous donation from the Courtois foundation, administered by the Fondation Institut Gériatrie Montréal at CIUSSS du Centre-Sud-de-l’île-de-Montréal and University of Montreal. The Courtois NeuroMod team is based at “Centre de Recherche de l’Institut Universitaire de Gériatrie de Montréal”, with several other institutions involved. See the cneuromod documentation for an up-to-date list of contributors (https://docs.cneuromod.ca).

Data from the Human Connectome Project was provided by the WU-Minn Consortium (Principal Investigators: David Van Essen and Kamil Ugurbil; 1U54MH091657) funded by the 16 NIH Institutes and Centers that support the NIH Blueprint for Neuroscience Research; and by the McDonnell Center for Systems Neuroscience at Washington University.

\clearpage
\newpage
\bibliographystyle{assets/plainnat}
\bibliography{refs}

\clearpage
\clearpage

\section{Methods}
\label{sec:methods}

\subsection{Approach overview}

Our objective is to predict the brain activity of participants exposed to naturalistic stimuli. This is framed as a high-dimensional regression task where the targets are the blood-oxygen-level-dependent (BOLD) signals detected by fMRI recording devices. We predict both at the cortical level, where the targets are the 20,484 vertices of the fsaverage5 surface~\citep{jenkinson2012fsl}, and at the subcortical level, where the targets are the 8,802 voxels of 8 subcortical regions defined by the Harvard-Oxford atlas~\citep{frazier2005structural,makris2006decreased,desikan2006automated}.

Our model takes as input three stimulus modalities (or as many as available, depending on the study): (i) the video clip being viewed by the participant, (ii) the audio being heard and (iii) the transcript of what is being heard. From these, we extract high-dimensional embeddings from the (frozen) intermediate layers of state-of-the art generative AI models along three modalities of interest: text, audio and video. 

This provides a timeseries of multimodal embeddings, which we feed to a (trainable) transformer to aggregate information across time. Each timestep is then projected to a low-dimensional latent space, from which a subject-conditioned layer predicts the brain responses at each selected area.

\subsection{Feature extraction}
\label{methods:features}

In this section, we describe our pipeline for feature extraction, which relies on modality-specific pretrained AI models. Note that these models are frozen, ensuring robustness to out-of-distribution data, as demonstrated by the results of TRIBE v1 on the Algonauts 2025 competition~\citep{d2025tribe}.

\paragraph{Text embeddings}
We extract "timed" text embeddings from the timestamped transcripts of the videos. For each word $w$ to embed, we prepend the preceding $k=1,024$ words in the transcript, which we feed through Llama-3.2-3B~\citep{grattafiori2024llama}. For each intermediate layer $l$, we extract the token(s) overlapping with the word $w$ and average them to obtain a contextualized word embedding of dimension $D_\text{text}=2048$. We then construct an evenly spaced grid at a frequency $f_{stim}=2$\,Hz, and for each time-bin, we sum the embeddings of words which overlap with the bin. 
This allows to temporally align the text features with the audio and video features. 

\paragraph{Audio embeddings}
To obtain audio embeddings, we extract audio files from the videos, split them into 60-second chunks, then feed these through Wav2Vec-Bert-2.0~\citep{chung2021w2v}. We then resample the hidden representations of the latter from 50\,Hz to $f_{stim}=2$\,Hz. For each intermediate layer $l$, this yields time series of embeddings of dimension $D_\text{audio}=1,024$.
Note that the resulting embeddings carry bidirectional information about both the past and future of the stimulus window, whereas text and video embeddings only contain information about the past.

\paragraph{Video embeddings}
For video embeddings, we again construct an evenly spaced grid at a frequency $f_{stim}=2$\,Hz, and for each bin of time, we feed 64 frames spanning the preceding 4 seconds to Video-JEPA-2-Giant~\citep{assran2025v}. For each intermediate layer $l$, we compress the tensor of activations by averaging over all patch tokens, yielding a time series of embeddings of size $D_\text{video}=1,280$. Note that this spatial averaging step was necessary to keep the size of the tensor manageable. However, it comes at the cost of discarding positional information, which we expect to deteriorate encoding performance in low-level visual areas which exhibit a retinotopic mapping~\citep{wandell2011imaging}.

\paragraph{Combining the modalities}

For each of the three modalities $m$, the feature extraction described above leads to a time series of embeddings at $f_{stim}=2$\,Hz, with embeddings of shape $[L_m,D_m]$, where $L_m$ and $D_m$ are the number of layers and dimensionality of the transformer of modality $m$. 
To compress these embeddings while retaining both low-level and high-level information, for each modality, we split the layers into $L$ groups, then average the tensor per group along the layer dimension, compressing to a shape $[L, D_m]$. 
We then concatenate the layers and feed the resulting vector through a linear layer with a shared output dimension $D=384$ followed by layer normalization. Finally, we concatenate the three modalities, leading to a time series of \textit{multimodal} embeddings of shape $D_{model}=3\times 384 = 1152$. This will be the input to our transformer encoder.

\paragraph{Implementation details}
We extract stimuli features from pretrained language, audio and video models available on the \texttt{HuggingFace} platform~\citep{jain2022hugging} and cache them as \texttt{Numpy} memmap arrays~\citep{harris2020array} for fast loading during the training of our encoding model. Feature extraction is completed in 24 hours on 128 V100 GPUs with 32GB of VRAM, and model training lasts 24 hours on a single such GPU. 


\subsection{Model}
\label{methods:model}

\paragraph{Transformer encoder}

We extract windows of duration $T=100$ seconds from these embedding time series, add learnable positional embeddings and a learnable subject embedding, then feed the result through a Transformer encoder\footnote{We use the transformer implementation from the \texttt{x-transformers} package available at \url{https://github.com/lucidrains/x-transformers}.} with 8 layers and 8 attention heads. This enables information to be exchanged between timesteps.
At the output of the transformer, we use an adaptive average pooling layer to decimate the embeddings from the stimulus frequency $f_{stim}=2$\,Hz to the resampled fMRI frequency $f_{fMRI}=1$\,Hz.

\paragraph{Modality dropout}

One desirable property of a multimodal encoding model is its ability to provide meaningful predictions in the absence of one or several modalities, for example for a silent movie or a podcast. To encourage this behaviour, while at the same time avoiding excessive reliance on one modality, we introduce modality dropout: during training, we randomly mask off each modality by zeroing out the corresponding input tensor with a probability $p=0.3$, resampling such that at least one modality is left unmasked.

\paragraph{Subject block}
TRIBE uses a subject block to handle subject specificities during training, inspired by~\citep{defossez2023decoding}. Specifically, we use a subject-conditional linear layer to project the transformer outputs to the $N_{target}$ encoding targets, which are either the vertices of the fsaverage5 cortical surface or the voxels of the subcortical regions. Denoting the number of subjects as $S$, the subject block can be viewed as a tensor of shape $(S, D_{model}, N_{targets})$ whose first dimension is indexed by the subject. This subject block can be fine-tuned on new subjects, as detailed in \cref{methods:training}. 

\paragraph{Unseen subject prediction}

For in-silico experimentation, one is generally interested in the group response to a particular stimulus. For this, we implement a form of "subject dropout": during training, with probability $p=0.1$, we bypass the subject block and instead feed the latents through a special "unseen subject" linear layer. This forces the model to maintain accurate predictions even without information on the subject, and enables our model to be used to predict group responses in a zero-shot manner.


\subsection{Training and validation}
\label{methods:training}

\paragraph{Training}

TRIBE is penalized via a mean-squared error term between the predicted and ground truth fMRI data, without any additional regularisation. We train it for up to 15 epochs with the AdamW optimizer~\citep{loshchilov2017decoupled} using a
batch size of 16. The learning rate is warmed up linearly to $10^{-4}$ over the first 10\% of steps, then decayed following a cosine learning rate schedule. We use early stopping based on the validation Pearson score, with a patience of 3 epochs. Model training can be performed under a day on a single V100 GPU with 32GB of VRAM thanks to the efficiency of the cached feature extraction, detailed in~\cref{methods:features}

\paragraph{Validation}

To avoid any form of data leakage, we ensure that the stimuli in the validation set have no overlap with the stimuli in the training set. For this, we either use the predefined splits defined by the authors, or define our own splits by randomly assigning 10\% of the podcasts/stories/movies to be held out. Our encoding metric is the Pearson correlation R. To compute it, we collect, for each subject and each parcel, the predicted and ground truth fMRI responses across all TRs of the validation set and compute the Pearson correlation. We then average the values over all subjects and parcels. We also refer to this metric as the “encoding score” of the model.

\paragraph{Fine-tuning}

Finetuning is then performed for a single epoch, with all parameters of TRIBE unfrozen, using the same hyperparameter configuration as described for training. Note that for fine-tuning, the subject block is retrained with $N_{subj}^{test}\neq N_{subj}^{train}$ given by the OOD study. To initialize the weights for the new subject block, the simplest option would be to use $L_{avg}$ for each test subject. However, since the number of subjects in the OOD studies is an order of magnitude larger than the training studies, the subject block obtained in this way (of shape $(S, D_{model}, N_{targets})$, with $N_{targets}=20,484$) can become prohibitively large. To circumvent this, we use low-rank matrix factorization: we factorize $L_{avg}$ as a product of low-rank matrices i.e. $ L_{avg}\simeq USV^T$, with a bottleneck rank $r=128$, using the SVD algorithm provided by $\texttt{torch.svd}$. The matrix $SV^T$ is then parametrized by a normal linear layer, while the matrix $U$, of shape $(r, N_{targets})$ becomes the new subject block.

\subsection{Linear baseline}

To provide a baseline for TRIBE, we built a linear multimodal brain encoder by simply replacing the transformer encoder by a linear convolution layer whose kernel spans 9 TRs and is offset by 5 seconds relative to the stimuli. The latter serves as a temporal receptive field to aggregate the signal across time. Note that this model is technically a deep linear model, since it contains a sequence of three linear operations: (i) the linear projection from the modality-specific embeddings to the shared mutimodal embedding space, (ii) the learnable convolutional layer and (iii) the final projection to cortical space. This linear model is trained in the same conditions as TRIBE, but with a maximum of 30 epochs instead of 15, as we observed that convergence is slower. We experiment both training this model across subjects, like TRIBE, and also training it separately on each subject. Interestingly, even this linear model benefits from multi-subject pretraining (\cref{fig:performance}).

\subsection{Statistics} 
\label{methods:stats}

To assess which cortical vertices are predicted above chance in \cref{fig:performance}, we follow~\cite{huth2016natural} and compare the estimated correlations to the null distribution of correlations between two independent Gaussian random vectors of the same length. Given the autocorrelation inherent to fMRI signals, we only keep one TR every 60 seconds to ensure the independence of the samples.

In \cref{fig:performance} and \cref{fig:ood}, we use paired $t$-tests with FDR correction across subjects to compare performances of different models.

\subsection{Datasets}
\label{methods:datasets}

In this section, we present the eight datasets that we curated from the literature for training and testing of our model: see \cref{tab:datasets_comparison} for a summary of their key features:
\begin{itemize}
    \item \textbf{Training:} Our model is trained jointly on four "deep" fMRI datasets where a small amount of participants were exposed to a large volume of naturalistic stimuli in varying conditions: silent videos~\citep{wen2018neural} (3 subjects, 35 hours), speech~\citep{lebel2023natural} (8 subjects, 85 hours), videos with sound but no speech~\citep{lahner2024modeling} (10 subjects, 61 hours) and multimodal videos~\citep{gifford2024algonauts} (4 subjects, 265 hours). 
    \item \textbf{Testing:} Our model is tested on four "broad" fMRI datasets where a large amount of participants were exposed to a small volume of naturalistic stimuli, two during speech listening~\cite{nastase2021narratives,li2022petit} (433 subjects, 326 hours) and two during movie watching~\cite{aliko2020naturalistic,van2013wu} (262 subjects, 338 hours). The choice to leave these for testing is motivated both by the observation that training on deep datasets leads to better performance~\citep{antonello2023scaling}, and by the fact that many subjects are required to obtain reliable estimates of the group-averaged response.
\end{itemize}

\paragraph{Courtois NeuroMod~\citep{gifford2024algonauts}}
The Courtois NeuroMod dataset consists of 3T recordings of six healthy human participants who watched the same naturalistic videos, namely the first six seasons of the popular TV series \textit{Friends} as well as four movies: \textit{The Bourne Supremacy, Hidden Figures, The Wolf of Wall Street} and \textit{Life} (a BBC Nature documentary). This amounts to an unprecedently large recording volume of over 80 hours of fMRI per subject. In the present work, we focus on a subset of four subjects curated for the Algonauts 2025 competition~\citep{gifford2024algonauts} (the other two subjects are not publicly available at the time of writing). We use the fMRI data preprocessed by the authors using fMRIPrep~\citep{esteban2019fmriprep}, co-registered to the MNI152NLin2009cAsym template.

\paragraph{Lebel2023~\citep{lebel2023natural}}
This dataset was recorded on a 3T machine while 8 healthy participants each listened to 27 complete, natural, narrative stories (370 minutes) from The Moth podcast over the course of five scanning sessions. Three of these participants also listened to a further 57 complete stories (629 minutes). We preprocess the data using fMRIPrep, co-registering to the MNI152NLin2009cAsym template.

\paragraph{BoldMoments~\citep{lahner2024modeling}}
This dataset was recorded on a 3T machine while 10 healthy participants watched 1,102 3-second video clips sampled from the Memento10k dataset. The clips were manually selected to encompass videos that contained movement (i.e., not static content), were filmed in a natural context, and represented a wide selection of possible events a human might witness. We use the train and test splits provided by the authors (train videos are viewed 3 times while test videos are viewed 10 times). We use the fMRI data preprocessed by the authors, co-registered to the MNI152NLin2009cAsym template.

\paragraph{Wen2017~\citep{wen2018neural}}
This dataset was recorded on a 3T machine while 3 healthy participants watched three hours of short video clips without sound. The clips were collected from Youtube and VideoBlocks and concatenated together into 8-minute continuous streams. Importantly, this is the only dataset considered without any audio content. We use the fMRI data preprocessed by the authors, co-registered to the MNI152NLin6Asym template.

\paragraph{Naturalistic NeuroImaging Database~\citep{aliko2020naturalistic}}
This dataset was recorded on a 3T machine while 86 healthy participants watched one of ten full-length movies. We use the fMRI data preprocessed by the authors, which is co-registered to the Colin27 template. 

\paragraph{Human Connectome Project~\citep{van2013wu}}
This dataset was recorded on a 7T machine as part of the Human Connectome Project, while 167 healthy participants watched a full movie. This dataset was selected for its large amount of participants and the 7T resolution, leading to a higher signal to noise ratio. We use the fMRI data preprocessed by the authors, co-registered to the MNI152NLin6Asym template.

\subsection{fMRI preprocessing}
\label{methods:preprocessing}

The datasets employed in this study feature diverse preprocessing pipelines and formats. Ideally, all datasets would be processed through an identical fMRIPrep workflow to yield time series in consistent volumetric and surface spaces (e.g., \textit{MNI152NLin2009cAsym} and \textit{fsaverage}). However, heterogeneities in data organization (e.g., non-BIDS compliance) and missing metadata rendered this approach unfeasible. We therefore adopted a harmonized workflow designed to minimize preprocessing variance and enhance model generalizability.

With the exception of the Lebel2023 dataset, all BOLD data were provided by the original authors already registered to a standard volumetric template (\textit{MNI152NLin2009cAsym}, \textit{MNI152NLin6Asym}, or \textit{Colin27}). For the Lebel2023 dataset, we performed preprocessing using fMRIPrep with default parameters, targeting the \textit{MNI152NLin2009cAsym} template.

\paragraph{Cortical extraction}

To map these volumetric data onto the \textit{fsaverage} surface, we utilized the following procedure. For each template $T$, we computed the geometric coordinates of the \textit{fsaverage} white and pial surfaces within the coordinate space of $T$ using FreeSurfer's \texttt{recon-all} and \texttt{surf-to-surf} tools. We then projected the volumetric time series onto these surfaces using \texttt{nilearn.surface.vol\_to\_surf}. Specifically, we employed the ``ball'' sampling method with a 3\,mm radius centered at a depth halfway between the pial and white matter boundaries.

\paragraph{Subcortical extraction}
To extract subcortical signals from the BOLD images, we mask them with the Harvard-Oxford atlas~\citep{frazier2005structural,makris2006decreased,desikan2006automated}, defined in MNI space at a resolution of 2 millimeters. This results in 8,802 voxels spanning 8 subcortical regions: hippocampus, lateral ventricles, amygdala, thalamus, caudate, putamen, pallidum and accumbens.

\paragraph{Rescaling and detrending}
The timeseries for each vertex were z-scored across each session. Importantly, detrending was then applied to the timeseries. This step was not included in the preprocessing of the Algonauts 2025 competition: in our experiments, neglecting this step led to confounds on datasets where slow drifts are particularly salient, such as Wen2017 and Lebel2023. Indeed, due to the long context window of our model, these drifts could be exploited by the encoding model to spuriously increase its encoding score in otherwise hard to predict brain areas. For this reason, detrending makes the encoding task performance significantly lower, but better reflects downstream encoding performance.

\paragraph{Resampling}
Finally, the signals are linearly resampled to $f_{fMRI}=1$\,Hz using $\texttt{np.interpolate}$ in order for the frequency to be consistent across the datasets, which use different repetition times.

\paragraph{Hemodynamic lag} There is a roughly 5-second delay between stimuli presentation and peak response in the brain. For this reason, we offset the fMRI timeseries by 5 seconds relative to the stimuli, meaning that to predict a window [0,T] of fMRI, TRIBE takes as input a window [-5, T-5] of stimuli.



\subsection{In-silico experiments}
\label{methods:in_silico}

In this section, we describe our protocol for replicating the IBC tasks \textit{in-silico}. Note that in all these experiments, TRIBE is used in unseen-subject mode (see~\cref{methods:model}).

\paragraph{Visual experiments}
We replicate the visual experiments from the \textit{FaceBody} and \textit{Visu} tasks of IBC, whose images are available at \url{https://github.com/individual-brain-charting/public_protocols}. All images are presented to TRIBE in a randomized order, for one second every eight seconds (transforming them to static videos). 
Note that the face image in~\cref{fig:in_silico_vision} is here used solely for illustrative purposes. The actual stimuli used for the study are available at \url{https://github.com/individual-brain-charting/public_protocols/tree/master/FaceBody/stimuli}.

\paragraph{Language experiments}
We replicate the experiments \textit{Bang}, \textit{Audio}, \textit{EmotionalPain} and \textit{RSVP} from the IBC datasets. 
For \textit{Bang}, since the original movie was not available for download, we simply contrast segments from the Algonauts dataset which contain speech versus those which do not. For \textit{Audio}, we display the audio segments in a randomized order, with 8 seconds between each segment.
For \textit{EmotionalPain} and \textit{RSVP}, where the stimuli consist in sentences, the replication is more involved since we need to obtain word timings. For this, we first translate the sentences from French to English, then feed the texts through a text-to-speech pipeline\footnote{https://github.com/kyutai-labs/pocket-tts} which yields an audio file. We then run a speech-to-text pipeline\footnote{https://github.com/m-bain/whisperX} to obtain word timings, which provides the expected dataframe of multimodal (text+audio) events. 

\paragraph{Contrast maps}
To obtain contrast maps for the visual experiments, we obtain contrast maps by simply selecting the predicted response at t=5 after the image is shown (which is the peak of the response as shown in \cref{fig:in_silico_vision}A), and substracting the average responses at t=5 for the other categories.

To obtain contrast maps for the lanugage experiments, we follow the protocol used for the IBC dataset. Specifically, we fit a General Linear Model (GLM), where the regressors are estimated through the convolution of TRIBE's predicted BOLD responses with the canonical Hemodynamic Response Function. We use the implementation provided by $\texttt{nilearn}$'s $\texttt{FirstLevelModel}$ with default parameters. 

\paragraph{Regions of interest}
To extract regions of interest, we used the Glasser Multimodal parcellation~\citep{glasser2016multi}. FFA, EBA, PPA, VWFA, A5, 45, STS, TPJ, MTG respectively correspond to the following ROI labels: FFC, V4t, PH, A5, 45, STSv, PGi, TE1a.



\subsection{Independent component analysis}
\label{methods:ica}

For this analysis, we use the "unseen" subject layer of TRIBE, a tensor of shape $(D_{model}, N_{targets})$ which maps from the latent space to each subject's cortical space.
To visualize the latent space, we apply Independent Component Analysis on this matrix, using the $\texttt{FastICA}$ implementation provided by $\texttt{sklearn}$ with default parameters and $n\_components=5$. Note that at odds with Principal Component Analysis, the order of the components does not carry any meaning for ICA.
The result is a collection of five vectors of size $N_{targets}$ which we can plot on the cortical surface.

To characterize these brain maps functionally, we compare them with five functional maps obtained by NeuroSynth~\citep{kent2026neurosynth} with the following keywords: "primary auditory", "language", "motion", "default network" and "visual". These maps are resampled from volumetric to cortical space via nilearn's $\texttt{vol\_to\_surf}$, then compared with the ICA components via Pearson correlation across all vertices of the cortical surface.

\clearpage
\newpage
\beginappendix
\section{Related works}
\label{app:related}

\paragraph{Limitations of current encoding models}
Motivated by the alignment between the representations learnt by AI models and those observed in the human brain~\citep{huth2016natural,schrimpf2018brain,caucheteux2022brains}, several teams have built encoding models to predict brain responses to natural stimuli from the activations of neural networks in response to images~\citep{yang2023memory,Adeli2023.08.02.551743,nguyen2023algonauts}, speech~\citep{millet2022toward} and text~\citep{toneva2019interpreting}. However, these encoding models are currently limited in three critical ways. 

First, \textit{linearity}: existing encoding approaches typically rely on ridge regression to map the AI model representations onto those of the brain. This assumes that these two sets of representations are linearly related, a phenomenon known not to be true~\citep{linsley2025can}.  Second, \textit{specificity}: due to large variability in brain responses across subjects, tasks and brain areas, existing encoding approaches typically train separate models for each subject, task and brain area, which prevents them from learning the co-occurring patterns which emerge in large heterogeneous datasets. Third, \textit{unimodality}: most existing encoding approaches predict brain responses from unimodal stimuli, which makes them incapable of capturing how the brain integrates information from multiple modalities~\citep{hu2025neural}. This is particularly limiting as it has been shown that cross-modal interactions occur not only in specific multisensory areas~\citep{gao2023audiovisual,beauchamp2005see}, but also in primary sensory areas~\citep{driver2008multisensory,stein2008multisensory}.

\paragraph{Deep encoders}
While there has been significant research on deep learning for multimodal brain decoding~\citep{dahan2025sim,scotti2024mindeye2,xia2024umbrae,zhou2024clip,kong2024toward}, the corresponding literature for brain encoding is more sparse. However, some recent works suggest to train recurrent models to predict brain responses from frozen visual or linguistic features~\citep{gucclu2017modeling, chehab2021deep}, or fine-tune existing pretrained models using the brain encoding objective~\cite{vattikonda2025brainwavlm}. Closest to our approach is~\citet{beliy2024wisdom}, which uses a transformer-based approach to predict the visual cortex. While these works relax the linearity assumption, they are restricted to a single sensory modality.

\paragraph{Multimodal encoders}
Conversely, a few studies have built encoding models on top of vision-language transformers, demonstrating gains compared to unimodal transformers~\citep{dong2023vision,oota2022visio,doerig2022semantic,wang2022incorporating,tang2023brain}. However, these works rely solely on linear mappings to model brain responses from the activations of the multimodal transformers. We believe this can be suboptimal for two reasons. First, multimodal transformers are still relatively new: at rare exceptions~\citep{jaegle2021perceiver,srivastava2024omnivec,abdin2024phi4}, they often only integrate static images and text (audio and video being significantly more compute-intensive), and tend to lag behind the performance of unimodal transformers. Second, and more fundamentally, the way these models integrate information across modalities may be very different from how the human brain does such multimodal integration. An ideal encoding pipeline should thus \textit{learn} how to best combine different modalities.

\paragraph{Comparison to TRIBE v1}

Motivated by the Algonauts 2025 brain encoding competition~\cite{gifford2024algonauts}, many teams have introduced multimodal brain encoding models, with some notable entries listed in~\cref{tab:leaderboard}. Among these, an earlier iteration of our current model, TRIBE v1, achieved the state-of-the-art. Compared to TRIBE v1, our model addresses the aforementioned shortcomings: (i) it is tasked to predict the 20k vertices of the fsaverage5 cortical map as well as subcortical regions (versus 1000 parcel activations from a cortical atlas for TRIBE v1); (ii) it is trained on 25 participants performing four different tasks (versus 4 participants on one task for TRIBE v1); (iii) it is equipped with an "unseen subject" module, which enables it to be tested on 695 participants coming from four held-out studies (TRIBE v1 was not tested on unseen subjects), (iv) thanks to the variety of tasks it was exposed to during training, TRIBE v2 generalizes to the non-naturalistic stimuli such as the controlled images of the IBC protocol~\cref{fig:in_silico_vision} and enables in-silico experimentation (for which TRIBE v1 was not tested).

\begin{table*}[h!]
    \centering
    \begin{tabular}{ccccccc}
    \toprule
    \textbf{Rank} & \textbf{Reference} & \textbf{Mean score} & \textbf{Subject 1} & \textbf{Subject 2} & \textbf{Subject 3} & \textbf{Subject 5} \\ 
    \midrule
    1 & Ours                  & \textbf{0.2146 $\pm$ 0.0312} & \textbf{0.2381} & \textbf{0.2105} & \textbf{0.2377} & 0.1720 \\
    2 & \citep{schad2025vibe}           & 0.2096 $\pm$ 0.0283 & 0.2353 & 0.2046 & 0.2268 & 0.1718 \\
    3 & \citep{eren2025multimodal}      & 0.2094 $\pm$ 0.0215 & 0.2233 & 0.2072 & 0.2271 & \textbf{0.1798} \\
    4 & \citep{villanueva2025predicting}& 0.2085 $\pm$ 0.0267 & 0.2295 & 0.2003 & 0.2300 & 0.1743 \\
    5 & Unpublished                    & 0.2055 $\pm$ 0.0291 & 0.2306 & 0.2010 & 0.2240 & 0.1662 \\
    \bottomrule
    \end{tabular}
    \caption{\textbf{\model{} achieves the first place in the Algonauts 2025 brain prediction competition 
    (Gifford et al 2024). We report the mean score ± standard deviation for the top five out of 263 teams.}
    }
    \label{tab:leaderboard}
\end{table*}

\begin{figure}
    \centering
    \includegraphics[width=\linewidth]{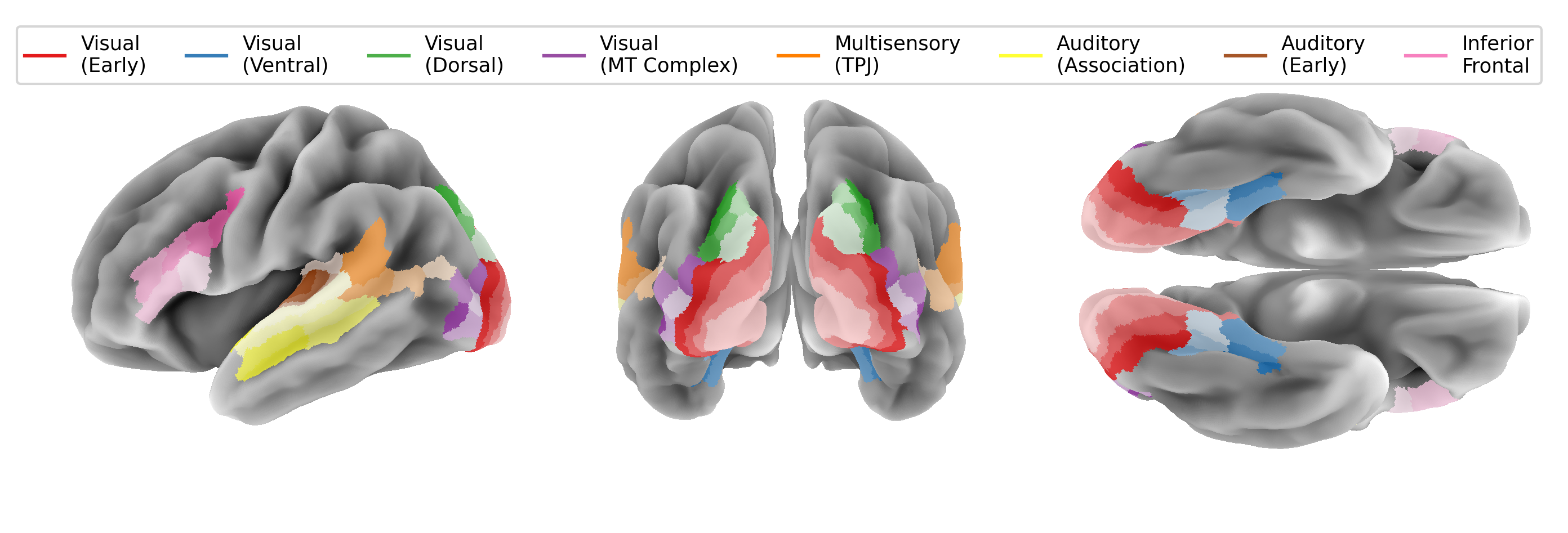}
    \includegraphics[width=\linewidth]{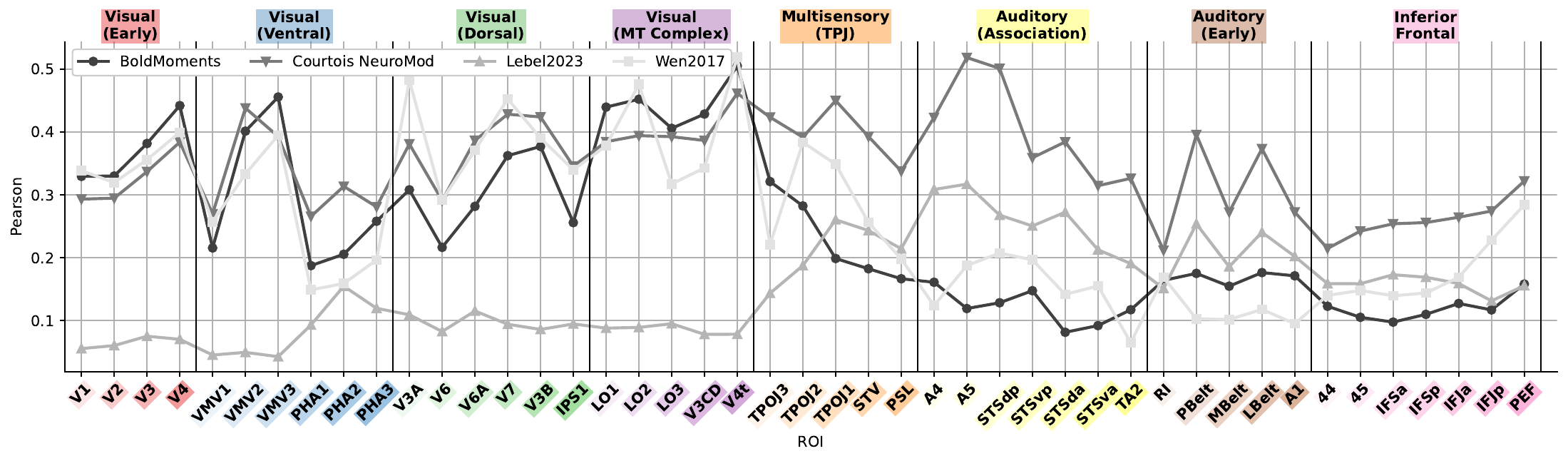}
    \caption{\textbf{Performance of TRIBE across cortical regions.}}
    \label{fig:rois}
\end{figure}

\begin{figure}
    \centering
    \includegraphics[width=.7\linewidth]{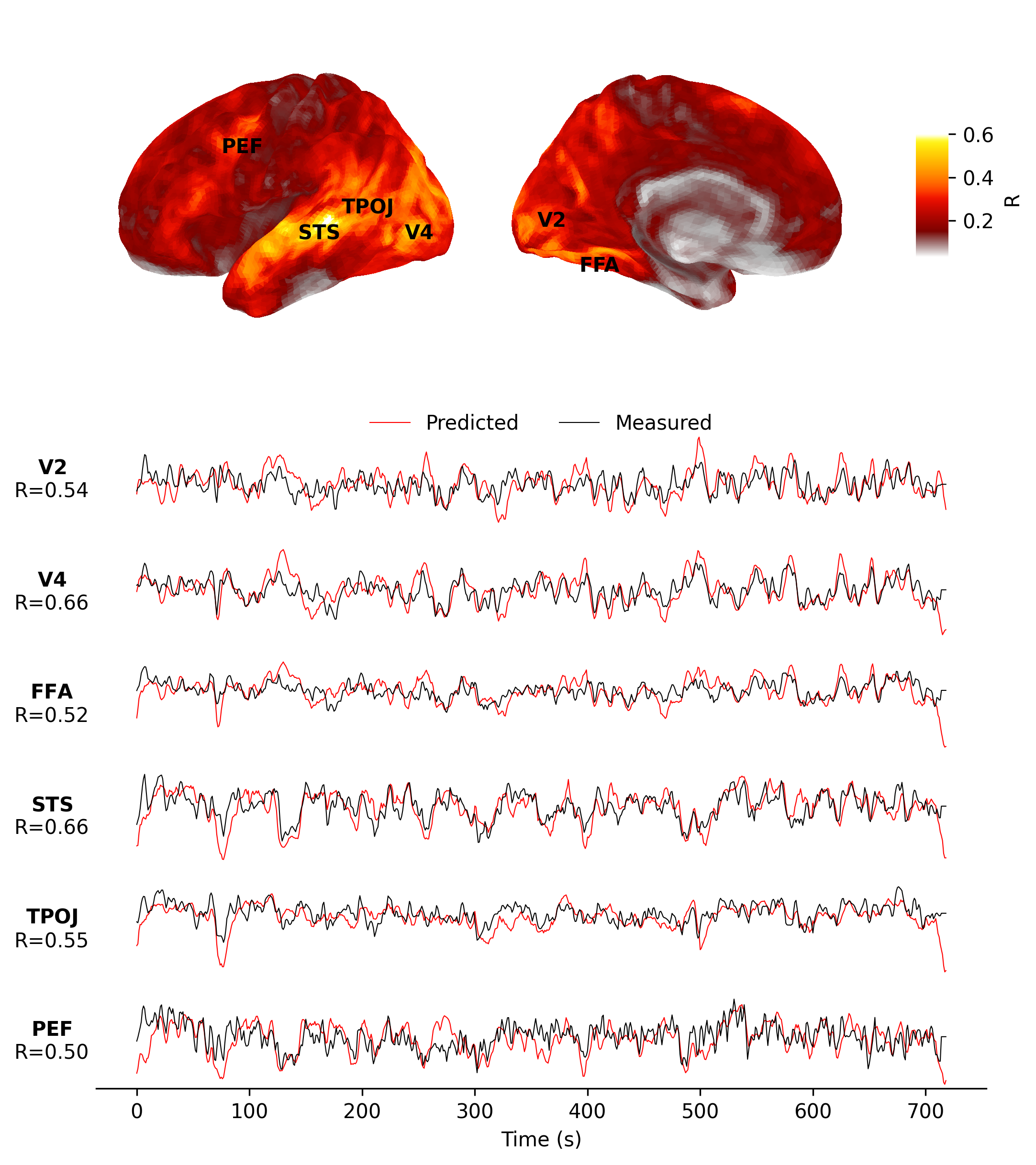}
    \caption{\textbf{Examples of decoded BOLD timeseries for various areas of the brain.}}
    \label{fig:timecourses}
\end{figure}

\begin{figure}
    \centering
    \includegraphics[width=\linewidth]{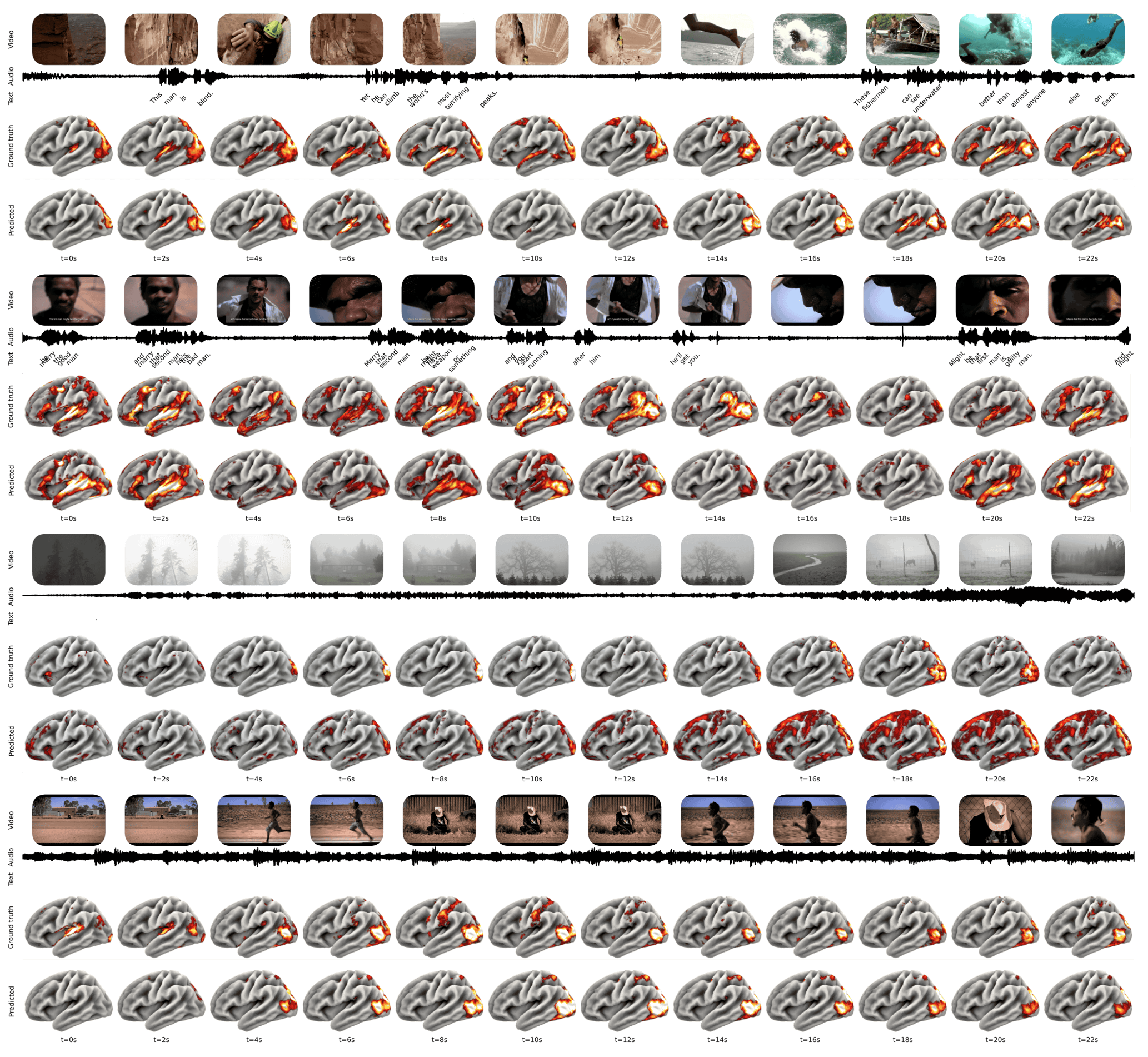}

    \caption{\textbf{Examples of video segments and the corresponding predicted and true brain responses.} In all cases, the brain activity is normalized to the 99\% percentile across the segment. See more examples in our interactive demo \url{https://aidemos.atmeta.com/tribev2/}.}
    \label{fig:demo}
\end{figure}

\newpage

\end{document}